\documentclass[sigconf]{acmart}

\usepackage{booktabs} 
\usepackage[utf8]{inputenc}
\usepackage{adjustbox}
\usepackage{graphicx}
\usepackage{subcaption}
\usepackage{multirow}
\usepackage{enumerate}
\usepackage{enumitem}
\usepackage{bbm}
\usepackage{amsmath}
\usepackage{amsfonts}
\usepackage{amssymb}
\usepackage{url}
\usepackage{acronym}
\usepackage{color}
\usepackage{mathtools}
\usepackage{tkz-graph}
\usepackage{tikz}
\usetikzlibrary{backgrounds}
\usetikzlibrary{trees,positioning,arrows}
\usepackage{wrapfig}
\usepackage[ruled,linesnumbered,lined,boxed]{algorithm2e}
\usepackage{array}

\newcommand{\RQ}[2]{
    \begin{description}[topsep=0pt,nosep=0pt, leftmargin=0.75cm, noitemsep,nolistsep]
    \phantomsection\label{section:setup:rq#1}
    \item[RQ#1] #2
    \end{description}
}

\newcommand{\RQRef}[1]{\textbf{\hyperref[section:setup:rq#1]{RQ#1}}}

\acmConference[CIKM'19]{The 28th ACM International Conference on Information and Knowledge Management}{November 03 - 07, 2019}{Beijing, China} 
\acmYear{2019}
\copyrightyear{2019}

\let\cal\mathcal
\begin{document}
\fancyhead{}

\title{Learning to Ask: Question-based Sequential \\Bayesian Product Search}


\author{Jie Zou}
\affiliation{%
  \institution{University of Amsterdam}
  \city{Amsterdam}
 \country{The Netherlands}}
\email{j.zou@uva.nl}

 \author{Evangelos Kanoulas}
 \affiliation{%
  \institution{University of Amsterdam}
  \city{Amsterdam}
  \country{The Netherlands}}
\email{e.kanoulas@uva.nl}

\renewcommand{\shortauthors}{J. Zou et al.}

\begin{abstract}
Product search is generally recognized as the first and foremost stage of online shopping and thus significant for users and retailers of e-commerce. Most of the traditional retrieval methods use some similarity functions to match the user's query and the document that describes a product, either directly or in a latent vector space. However, user queries are often too general to capture the minute details of the specific product that a user is looking for. In this paper, we propose a novel interactive method to effectively locate the best matching product. The method is based on the assumption that there is a set of candidate questions for each product to be asked. In this work, we instantiate this candidate set by making the hypothesis that products can be discriminated by the entities that appear in the documents associated with them. We propose a Question-based Sequential Bayesian Product Search method, QSBPS, which directly queries users on the expected presence of entities in the relevant product documents. The method learns the product relevance as well as the reward of the potential questions to be asked to the user by being trained on the search history and purchase behavior of a specific user together with that of other users. The experimental results show that the proposed method can greatly improve the performance of product search compared to the state-of-the-art baselines. 
\end{abstract}


\maketitle

\section{Introduction}

Purchasing goods and services over the Internet is by many considered a convenient and cost-effective shopping method. This has led to an ever-increasing focus on online shopping technologies. The first and foremost stage of online shopping is generally recognized to be that of product search ~\cite{rowley2000product}. In product search, the users usually formulate queries to express their needs and find products of interest by exploring the retrieved results. %

When it comes to product search, the majority of methods use some similarity functions to match the query and documents that describes a product. In these cases, the query and product documents \footnote{In this paper, we use product documents to refer to product descriptions and reviews.} are usually represented as vectors in observed or latent vector space and the distance between these vectors is computed ~\cite{van2016learning, ai2017learning, guo2018multi, guo2019attentive}. However, the queries in product search are often too general to capture the minute details of the exact product a user is looking for~\cite{duan2013supporting, duan2013probabilistic}.

Although the problem of efficiently finding the best match for a query with similarity and representation learning in a given set has been studied before~\cite{van2016learning, ai2017learning, guo2018multi, guo2019attentive}, interactive methods that can help users better specify their needs still remain underexplored. The main focus of this work is to effectively find the best matching product for the user by asking "yes"/"no" questions to the searchers. 
Given a pool of available questions to be asked on products, for which the answer can either be ``yes'', or ``no'', or ``not sure'', our method performs a duet training, to learn both the product relevance and the informativeness of a question.
That is, our proposed method uses limited data from existing users to construct a system belief over product relevance for any new user/query. Further, it is trained to select questions with the highest potential reward based on the system belief over product relevance. In particular, our method extends Generalized Binary Search (GBS) ~\cite{nowak2008generalized} over questions to find the question that best splits the probability mass of predicted product relevance and has the highest trained reward for the remaining of the products. After the question is being asked and answered by the user a posterior belief over product relevance is obtained and used for the selection of the next question. Lastly, the true belief over product relevance for the user is revealed sequentially through interactions with the user to generate the final product search results.

The main contribution of this paper is three-fold: 
\begin{itemize}
\item A novel interactive product search method based on constructed questions, QSBPS, which directly queries users about the expected presence of an informative term in product related documents.
\item A method that learns question reward and cross-user system belief with limited data.
\item An extensive analysis of the performance of the algorithms that includes an analysis of noise tolerance in user answers.
\end{itemize}

The evaluation results show that our approach can achieve better results compared to the state-of-the-art baselines.

\section{Related Work}
\label{sec:relwork}

\subsection{Product Search}
E-shopping and e-retailing are attracting more and more attention, which has led to new developments in technology ~\cite{rowley2000product}. Most previous product search methods focus on structured information about products, e.g., brands, types and categories. Based on a semantically annotated product family ontology, ~\citet{lim2010multi} presented a multi-facet product search and retrieval framework. ~\citet{vandic2012faceted, vandic2013facet} noticed that product information on pages is usually not well-structured, and proposed a faceted product search algorithm by using semantic web technology. ~\citet{duan2013supporting, duan2013probabilistic} observed that there is a vocabulary gap between product specifications and search queries, and developed a probabilistic mixture model to systematically mine product search logs by learning an attribute-level language model. %
~\citet{van2016learning}, on the other hand, introduced a latent vector space model to learn representations for products based on their associations with unstructured documents, which avoids the limitations of searching in structured data. They learned distributed representations of a given word sequence and products as well as a mapping between query and product space. ~\citet{ai2017learning} presented a hierarchical embedding model to learn semantic representations for queries, products, and users. They constructed a latent space retrieval model for personalized product search. ~\citet{zamani2018joint} proposed a general framework that jointly models and optimizes a retrieval model and a recommendation model. Some studies also investigated the effectiveness of incorporating external information ~\cite{guo2018multi}. %
~\citet{guo2018multi} presented the translation-based search (TranSearch) model, which tries to match the user's target product from both textual and visual modalities by leveraging the "also\_view" and "buy\_after\_viewing" of products. Then ~\citet{guo2019attentive} proposed an Attentive Long Short-Term Preference model (ALSTP) for personalized product search by considering the long-term and short-term user preferences using two attention networks.
Similar to previous work we also hypothesize that there is a gap between the language of product documents and user queries. Different from past work, we propose a question-based sequential Bayesian interactive learning on user preferences to learn their actual needs. Past work is complementary to ours, since it can be leveraged to inform a prior on product relevance, but it can also be used to construct questions on the basis of product structured information.
  
\citet{zhang2018towards} proposed a unified paradigm for product search and recommendation, which constructs questions on aspect-value pairs, to ask the user questions over aspects. Their model obtains the opinion of the user (i.e., value of the aspect-value pair) for the "aspect'" as feedback and thus expand the representation of the user query. Different from that work based on aspect-value pairs and constructs the question by manual predefined language patterns, we ask questions about automatically extracted informative terms without complex language patterns. Further, the selection of questions in \citet{zhang2018towards} relies on the log-likelihood of probability estimation over limited aspects while our question selection is based on the cross-user duet learning of question effectiveness and user preferences.

\subsection{Interactive Search}

Quantifying relevance on the basis of users' queries, or learning a model of relevance from past queries, cannot always capture the minute details of relevance, not only in product search, but also in other search tasks, such as those that require high recall ~\cite{cormack2014evaluation}. 
Interactive information retrieval, instead, suggests putting the human in the loop and learning a relevance model throughout an interactive search process, where users provide feedback on the relevance of presented documents, and the model adapts to this feedback ~\cite{cormack2015autonomy}. 
 Most of the methods take a special treatment of the query ~\cite{zhang2018towards}, %
typically expanding it with terms from labeled documents. However, query expansion has show suboptimal performance ~\cite{cormack2014evaluation}, in part because handling the relation between the original query and feedback documents is challenging ~\cite{lv2009adaptive}. Active learning~\cite{cormack2015autonomy}
 and multi-armed bandits ~\cite{hofmann2013balancing} 
have also been proposed to iteratively learn task-specific models. %
Different from the afore-mentioned methods that focus on receiving feedback at the level of documents, our interactive method asks explicit questions to the users in terms of entities contained in the documents of the collection.
Similar to our work, ~\citet{Wen:2013} proposed a sequential Bayesian search (SBS) algorithm for solving the problem of efficiently asking questions in an interactive search setup. They learn a policy that finds items in a collection using the minimum number of queries. Based on ~\citet{Wen:2013}, ~\citet{zou2018technology} devised an SBSTAR algorithm to find the last few missing relevant documents in Technology Assisted Reviews by asking "yes" or "no" questions to reviewers. Our QSBPS algorithm differs by performing a cross-user duet training, to learn not only a belief over product relevance but also the reward over the performance of questions, as well as their noise-tolerance.

\subsection{Learning to Ask}
  Learning to Ask is a new field of study that has recently attracted considerable attention. A number of studies focus on identifying good questions to be asked in a 20 Questions game setup. ~\citet{chen2018learning} presented a Learning to Ask framework, within which the agent learns smart questioning strategies for both information seeking and knowledge acquisition.%
 ~\citet{hu2018playing} proposed a policy-based reinforcement learning method also within a 20 Questions game setup, which enables the questioning agent to learn the optimal policy of question selection through continuous interactions with the users. Both aforementioned works introduce data-hungry techniques, which require having large numbers of repeated interactions between the users and the information seeking system to train upon. Different from these approaches, our method does not require having multiple searchers and their interactions for a given product.

\begin{figure}[t]
  \caption{Research framework}
  \includegraphics[width=1.0\columnwidth]{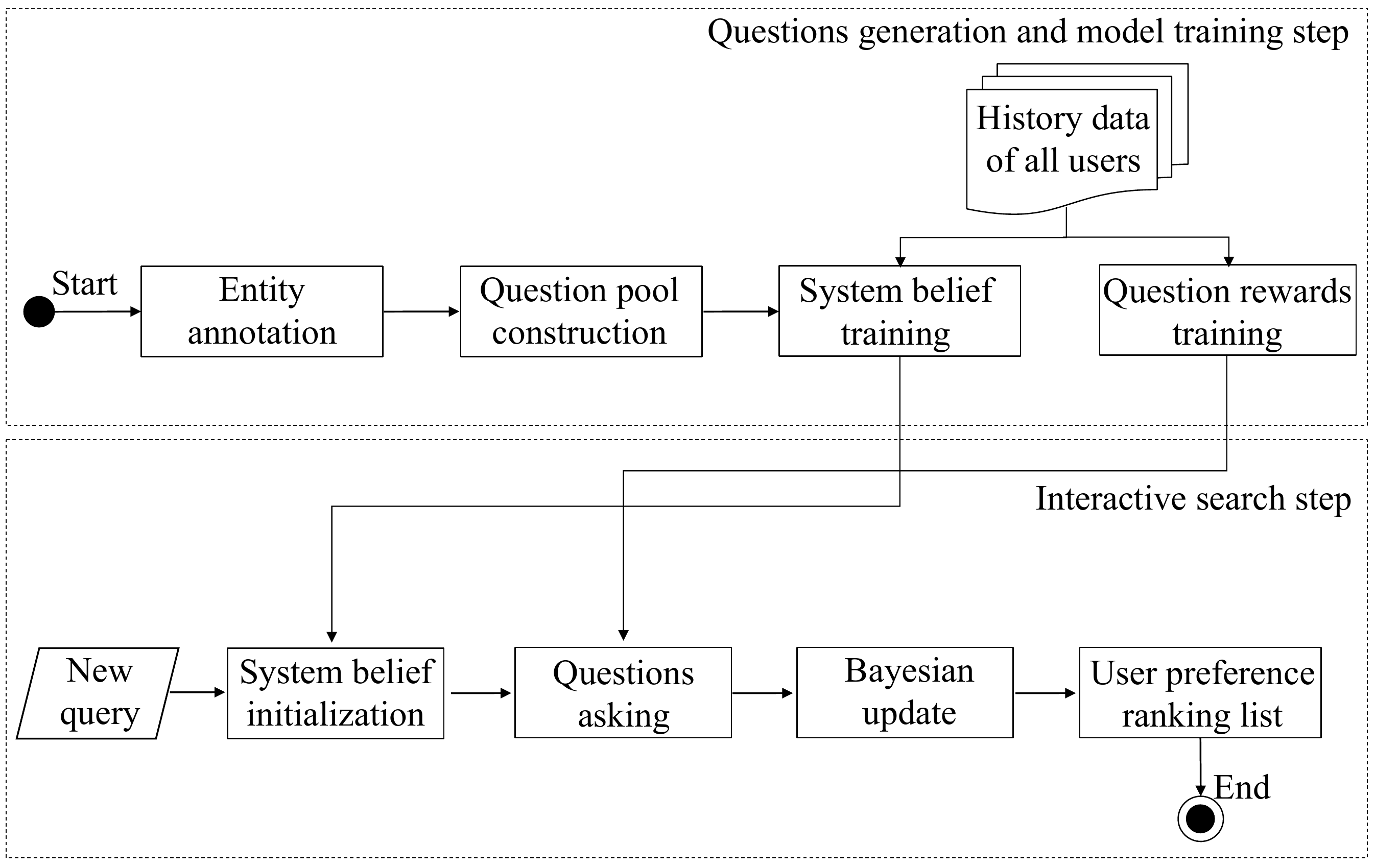}
  \label{fig:framework}
\end{figure}

\section{Methodology}
\label{sec:meth}
In this section, we provide a detailed description of the proposed method \footnote{source code: \url{https://github.com/UvA-HuMIL/QSBPS}}. The research framework is shown in Figure~\ref{fig:framework}. Our approach consists of three parts: (a) the construction of a pool of questions; (b) the system belief and the question reward training using historical data from each user individually, and across users; and (c) the interactive search step, which sequentially selects questions to be asked to the user and updates the user preferences. The focus of this work lies in the two latter parts. 

\subsection{Question Pool Construction}

The proposed method of learning informative questions to ask to users, described in details in Sections~\ref{subsec:cro} and \ref{sec:QSBPS}, depends on the availability of a pool of questions on product properties. That is, given a product, the user should be able to answer the question, with a reference to the relevant product, with a ``yes'' or a ``no''. 

There are different methods one could employ to construct such a set of questions. For instance, one can use labelled topics ~\cite{ZOU201719, JieZOU20162016EDP7052, zou2016duplication}, extracted keywords ~\cite{gupta2017keyword}, 
item categories and attributes, or extract triplets and generate a rich set of questions based on these triplets ~\cite{reddy2017generating}. In this work, we take a rudimentary approach. Our assumption is that a user can discriminate between products based on the language of the documents (i.e., descriptions and reviews) they are associated with.
We then identify informative terms (instantiated by entities in this work) using the entity linking algorithm TAGME ~\cite{Ferragina:2010}, similar to previous research ~\cite{xiong2015esdrank,xiong2017word}. 
 We assume that these informative terms comprise the most important characteristics of a product, and we generate questions about the presence or absence of such entity in the product related documents. That is, we instantiate the question candidate set by identifying entities in the product related documents. For example, from the following description of a product, "Apple iPhone XS (64GB), gold color. The most durable glass ever in a smartphone. And a breakthrough dual-camera system. iPhone XS is everything you love about iPhone. Taken to the extreme.", the extracted entities can be "Apple", "iPhone XS", "gold color", "smartphone", "dual-camera system", and "iPhone".
We don't use any filter on the annotation scores of the TAGME results, i.e.\ all annotations are being considered, which is also a widely used setting in previous work ~\cite{xiong2015esdrank, raviv2016document}. 
 After that, the proposed algorithm asks a sequence of questions of the form “Are you interested in [entity]?” to locate the target product. 

\begin{algorithm}[tb]
\SetKwInOut{Input}{input}
\SetKwInOut{Output}{output}
\Input{A product documents collection, ${\cal D}$, the set of annotated entities in ${\cal D}$, ${\cal E}$ , a set of topics $\mathcal{T}$, a prior Dirichlet parameter, $\mathbf{\alpha_0}$}
\Output{System belief $\mathbb{P}_{t}(\pi)$, question rewards $R_{t}(e)$}

\ForEach{topic $t \in {\cal T}$}{
  Compute the initial user preference with $\mathbf{\alpha_0}$: $\pi^*_0(d) = \mathbb{E}_{\pi \sim \mathbb{P}_0}[\pi(d)] ~ \forall d \in {\cal D}$ \
  
  Let $\mathcal{D}_t$ be the set of products within $t$\
  
  $n \leftarrow 0$\
  
  \ForEach{$d \in \mathcal{D}_t$}{
    
    \ForEach{entity $e \in {\cal E}$}{
      Update the system's belief $\mathbb{P}$ using Bayes' rule: $\mathbb{P}_{n+1}(\pi) \propto \pi(d)\mathbb{P}_{n}(\pi) ~ \forall \pi$ \
      
      Calculating reward for each entity: \
      $R_d(e)= \frac{ I_{before} - I_{after}}{|U|} $ \

      $n \leftarrow n + 1$\
    }
  }
  
  Output trained system belief $\mathbb{P}_{t}(\pi) = \mathbb{P}_{n+1}(\pi)$\
  
  Output average reward for each entity: $R_{t}(e)= \frac{1}{|\mathcal{D}_t|} \sum_{d \in \mathcal{D}_t} {R_d(e)}$\
  
}
\caption{QSBPS Offline Learning}
\label{algo_train}
\end{algorithm}

\begin{algorithm}[tb]
\SetKwInOut{Input}{input}
\SetKwInOut{Output}{output}
\Input{A product documents collection, ${\cal D}$, the set of annotated entities in ${\cal D}$, ${\cal E}$, a set of topics $\mathcal{T}$, a number of questions to be asked, $N_q$, the system belief $\mathbb{P}_{t}(\pi) \forall t \in \mathcal{T}$, and the question rewards $R_t(e) \forall t \in \mathcal{T}$}
\Output{User preference $\pi^*_{N_q}$}
\ForEach{topic $t \in \mathcal{T}$ }{

  Load $\mathbb{P}_{t}(\pi)$, $R_t(e)$\
  
  $l \leftarrow 1$\
  
  System belief initialization: $\mathbb{P}_{l}(\pi) \leftarrow \mathbb{P}_{t}(\pi)$\
  
  \While{$l \le N_q$ and $|U_l| > 1 $}{
    Compute the user preference with $\mathbb{P}_{l}(\pi)$: 
    $\pi^*_l(d) = \mathbb{E}_{\pi \sim \mathbb{P}_l}[\pi(d)] ~ \forall d \in {\cal D}$ \ 
    
    Find the optimal target entity: $e_l =$     
    $\arg\min_{e}|\sum_{d \in u_l} (2 \mathbbm{1}{\{e(d)=1\}} - 1)\pi^*(d) | -  \gamma * R_t(e)$\
    
    Ask the question about $e_l$, observe the reply $e_l(d^*)$\    

    Remove $e_l$ from entity pool\    
    
    $U_{l+1} = U_l \cap {i  \in {\cal D} : e_l(i)=e_l(d^*)}$\

    Update the system's belief: $\mathbb{P}_{l+1}(\pi) \propto \pi(d)\mathbb{P}_l(\pi) ~ \forall \pi$\
    
    $l \leftarrow l + 1$\
  }
}
\caption{QSBPS Online Learning}
\label{algo1}
\end{algorithm}

\subsection{Cross-user Duet Learning}
\label{subsec:cro}
In this section, we describe the training algorithm that jointly learns (a) a belief over the effectiveness of questions (entities) in identifying relevant products, $R_t(e)$, and (b) a system belief over the relevance of the products, $\mathbb{P}_{l}(\pi)$.
Instead of using the SBS algorithm ~\cite{Wen:2013}, a data-hungry algorithm that requires a large amount of training data for each user's request, our approach uses a duet learning approach on the given topic, $t$.\footnote{Topics in this paper are product subcategories, which can represent user queries.} The proposed method updates the system belief over relevance and the entity effectiveness after every question is being answered by the user, performing well using limited and weak signals. The system belief training over products learns the interest of the users over products, while the entity effectiveness training over entities learns the reward or informativeness of questions, and thus finds the optimal policy for asking questions. 
Our algorithm performs two rounds. During the offline phase, the algorithm learns what is the average users' preference over the products within each product category, and how effective are entities in identifying these products. Then, during the online interactive search phase, the algorithm continues learning product relevance on the basis of the user's answers to the algorithm's questions. %

The offline training phase is described in Algorithm ~\ref{algo_train}. We assume that there is a target relevant product $d^* \in \mathcal{D}$  \footnote{In the rest of paper, "product" and "product document" will be used interchangeably}. The user preferences for the products are modelled by a (multinomial) probability distribution $\pi^*$ over products $\mathcal{D}$, and the target product is drawn i.i.d. from this distribution. We also assume that there is a prior belief $\mathbb{P}_0$ over the user preferences $\pi^*$, which is a probability density function over all the possible realizations of $\pi^*$. The prior system belief $\mathbb{P}_0$ is a Dirichlet distribution, with a hyper-parameter $\alpha_0$, which can be set by using any other product search algorithm that measures the lexical or semantic similarity between the query and product documents, or any collaborative filtering method. In this paper, we use an uninformative uniform system belief distribution by setting all $\alpha_0$'s to 1 to isolate the effect of the proposed method. That is, the user preference is initialized to be the same for each product.

During the duet training phase, there is a training set $D_{t}$ for each topic, which contains all of the training products for this topic. For each product in $D_{t}$, we generate a set of questions based on all the entities in the collection, and we obtain a posterior belief using the Bayes' rule after every question for the target training product $d$ is being answered, and calculate the reward $R(e)$ for each question (entity). We then get the average reward for each entity to be used in the online interactive search, and obtain the trained system belief $\mathbb{P}_{t}(\pi)$, which is also used as a prior belief over products during the interactive search.
  
\textbf{System Belief:} We learn the system belief from the training data of all users on a certain topic, assuming that the training products on a certain topic are related in the entity embedding space, and thus can provide useful guidance. One could also learn user personalized preferences and entity informativeness, however, we do not do so, hypothesizing that users can buy significantly different products.

Let $\mathbb{P}_l$ be the system’s belief over $\pi^*$ in the $n$-th question. We compute the user preference $\pi^*_l(d)$ in the $n$-th question by,
\begin{equation}
  \pi^*_n(d) = \mathbb{E}_{\pi \sim \mathbb{P}_n} [\pi(d)]\forall d \in \mathcal{D}
\end{equation}

Similar to ~\citet{Wen:2013}, we model the user preference $\pi^*$ by a multinomial distribution over products $\mathcal{D}$. Then, the system updates its belief after observing the user answer to a question asked, which is sampled i.i.d. from $\pi^*$. 
\begin{equation}
  \mathbb{P}_{n+1}(\pi) \propto \pi(d)\mathbb{P}_n(\pi) ~ \forall \pi
\end{equation}

We model the prior $\mathbb{P}_0$, by the conjugate prior of the multinomial distribution, i.e., the Dirichlet distribution, with parameter $\alpha$. Further, we define the indicator vector $Z_l(d) = \mathbbm{1}\{e_l(d) = e_l(d^*)\}$, where  $e_l(d)$ means whether the product $d$ contains entity $e_l$ or not. 
From Bayes' rule, the posterior belief prior to the question $l$ is: 
\begin{equation}
  \label{eq:posterior}
  \mathbb{P}_{n+1} = Dir(\alpha + \sum_{j = 0}^{n} Z_j)
\end{equation}

From the properties of the Dirichlet distribution, we have:
\begin{equation}
  \label{eq:pref}
  \pi^*_n(d) = \mathbb{E}_{\pi \sim \mathbb{P}_n} [\pi(d)] = \frac{\alpha(d)+ \sum_{j=0}^{n} Z_j(d)}{\sum_{d' \in \mathcal{D}} (\alpha(d')+ \sum_{j=0}^{n} Z_j(d'))}
\end{equation}

where $\alpha(d)$ is the $i_{th}$ entry of $\alpha$, which corresponds to product $d$. Therefore, the user preference $\pi^*_l$ can be updated by counting and re-normalization.

\textbf{Question Reward:} For the question reward learning, we use historical training data to learn the reward of each entity. We define the following simple reward function, which can learn the ranking rising ratio of the target product relative to the candidate products version space when training. 

\begin{equation}
R(e)= \frac{ I_{\text{before}} - I_{\text{after}}}{|U|} 
\label{equ3}
\end{equation}

where the ${I_{\text{before}}}$ is the index of the target product in the ranked list before asking the question about entity $e$, the ${I_{\text{after}}}$ is the index of target product in the ranked list after asking the question about entity $e$, and $|U|$ is the number of products in the candidate set $U$. The ranking list is generated according to the user preference $\pi^*_{l}$ over the products. Note that we use the worst ranking index as the index of product in the ranked list when there are ties over the user preferences. Thus, in the first question, ${I_{\text{before}}}$ is initialized to the last ranking index, which is equal to the number of products in the collection.

After the system belief learning and question reward learning, the model uses its current user preference $\pi^*(d)$ from the belief $\mathbb{P}_t$ and the estimated reward $R_t(e)$ to derive a policy to find the optimal entity to query. 

\subsection{QSBPS Algorithm}
\label{sec:QSBPS}
In this section, we introduce two versions of the QSBPS algorithm. The first version assumes that there is no noise in the answers of a user. That is, when an entity appears in the text of the relevant product the user  gives a correct positive answer, while when it does not the user gives a correct negative answer (see user simulation in Section ~\ref{UserSimu}). The online interactive learning of our proposed algorithm is provided in Algorithm ~\ref{algo1}. 
We first load in the trained system belief $\mathbb{P}_{t}(\pi)$ and question rewards $R_t(e)$, then compute the user preference with prior belief equal to $\mathbb{P}_{t}(\pi)$, and find the optimal entity $e_l$ by Equation ~\ref{equ4}. Inspired by ~\citet{Wen:2013}, we extend GBS over entities to find the entity that best splits the probability mass of predicted product relevance closest to two halves, but also maximize the question reward for the remaining of the products during $l_{\text{th}}$ question, as the optimal entity. 
\begin{equation}
e_l = \arg\min_{e}\Big|\sum_{d \in u_l} (2 \mathbbm{1}\{e(d)=1\} - 1)\pi^*(d) \Big| -  \gamma * R(e)
\label{equ4}
\end{equation}
where $e_l$ is the $l_{\text{th}}$ choosen entity, $u_l$ is the set of products of the candidate version space when asking the $l_{th}$ question, $e(d)$ expresses whether the product $d$ contains the entity $e$ or not, while $\gamma$ is the weight to trade the question reward $R(e)$. We ask whether the entity $e_l$ is present in the target product that the user wants to find, $d^*$, observe the reply $e_l(d^*)$, and remove $e_l$ from the entity pool. Then we reduce $U_l$, update the system's belief $\mathbb{P}_l$ using Bayes' rule and recalculate the user preference, i.e., the user preference is updated sequentially by Bayesian learning that refers to sequential Bayesian based search. Since the user preference $\pi^*$ is a multinomial distribution over products $\mathcal{D}$, and $\mathbb{P}_t$, a Dirichlet distribution, updating the system belief is performed in a similar to Eq.\ref{eq:posterior} manner.

In Algorithm ~\ref{algo1}, we make the assumption, that users, when presented with an entity, know with 100\% confidence whether the entity appears in the target product. To relax this assumption we also propose a noise-tolerant version of the algorithm. That is, we allow the user to make mistakes and provide the algorithm with the wrong answer. We integrate the probability that the user will give the wrong answer to a question about entity $e$, $h(e)$, into the new objective function, at line 7 of Algorithm ~\ref{algo1},

\begin{equation}
e_l = \arg\min_{e}\Big|\sum_{d \in {\cal D}} (2 \mathbbm{1}{\{e(d) = 1 \}}  - 1 ) \pi^*(d) \Big| + 2\beta*h(e) -  \gamma * R_l
\end{equation}

We observe the noisy answer and update the posterior system belief according to this noisy answer. Intuitively, a question will be chosen to be asked not only if it is about an informative entity, but also if this entity is the one for which users have a good confidence in providing an answer. The experiments will be discussed in \RQRef{4}. Regarding the error rate for each entity $h(e)$, we consider two settings: In the first setting all of the $h(e)$ are simply set to equal values, and we experiment with different error rates that range from 0.1 to 0.5 with a step of 0.1, to explore the performance trend of our model with different error rates. An error rate $h(e)$ of 0.5 means that the user has a 50\% probability to give the wrong answer. In the second setting, given that users are usually more confident in their answers about an entity $e$ if $e$ is frequently occurring in the given topic, we define $h(e)$ as a function of average term frequency (TF) in the topic for each entity, which is in the range of (0, 0.5]:

\begin{equation}
\label{equ:errorrate}
h(e)=\frac{1}{2(1+\text{TF}_{\text{avg}}(e))}
\end{equation}

Where $\text{TF}_{\text{avg}}(e)$ represents the average term frequency of entity $e$ in the given topic. The choice of this function is ad-hoc and any other function of any other characteristic of entities could also be used. Ideally, one should conduct a user study to identify a reasonable error rate function, the properties of entities that affect the error rate, or even the characteristics of the users that influence the error rate. We leave such a user study as future work.

\section{Experiments and Analysis}
\label{sec:exp}
Through the experiments conducted in this work we aim to answer the following research questions:
\RQ{1}{What is the impact of the number of questions asked and the parameter $\gamma$ that trades the weight of question reward?}
\RQ{2}{What is the influence of using user reviews along with product descriptions?}
\RQ{3}{Does our duet training by using other users' data help?}
\RQ{4}{What is the performance when considering noisy answers?}
\RQ{5}{How effective is our proposed method for finding the best matching product compared to prior works?}

\subsection{Experimental Setup}
\label{ExpSetup}

\subsubsection{Dataset.} We use the collection of Amazon products \footnote{http://jmcauley.ucsd.edu/data/amazon/} ~\cite{mcauley2015inferring}. 
 Similar to ~\citet{van2016learning}, we use the same four different product domains from the Amazon product dataset, but due to the limited space, we only report two domains in this paper, which are "Home \& Kitchen", and the "Clothing, Shoes \& Jewelry".
Statistics on these two domains are shown in Table ~\ref{table:data}. The documents associated with every product consist of the product description and the reviews provided by Amazon customers. To construct topics (queries) we use the method employed in ~\citet{van2016learning}, and ~\citet{ai2017learning}, i.e.\ we use a subcategory title from the above two product domains to form a topic string.
Each topic (i.e. subcategory) contains multiple relevant products and products can be relevant to multiple topics.
After that, we remove the topics which contain just a single product, since having a single relevant product does not allow constructing a training set and test set. Similar to ~\cite{ai2017learning}, we randomly split the dataset to 70\%  and 30\% subsets, and we use 70\% of the products for each topic (i.e. subcategory) for training. We also use a 10\% of the data as validation set. The validation set is used to select the optimal parameters to avoid overfitting. %

\begin{table}
\captionsetup{font={small}}
\caption{Overview of the dataset. M denotes Metadata only and M\&R denotes Metadata \& Reviews. Arithmetic mean and standard deviation are indicated wherever applicable.}
\label{table:data}
\centering
\small
\begin{tabular}{p{0.41\columnwidth}p{0.23\columnwidth}<{\centering}p{0.24\columnwidth}<{\centering}}
\hline
& Home \& Kitchen & Clothing, Shoes \& Jewelry \\
\hline 
Number of products & 8,192 & 16,384 \\
Number of description docs & 8,192 & 16,384 \\
Number of reviews & 79,938 & 77,640 \\
Length of documents & 70.02 $\pm$ 73.82 & 58.41 $\pm$ 61.90  \\ 
Reviews per product & 9.76 $\pm$ 52.01 & 4.74 $\pm$ 18.60 \\
Number of topics & 729 & 833\\
Products per topic & 11.24 $\pm$ 31.16 & 19.67 $\pm$ 55.24\\
Number of entities (M)  & 232,086    & 385,727  \\
Number of entities (M\&R) & 1,483,659  & 1,408,828 \\
Entities per product (M) &   28.33 $\pm$ 23.93 &   23.54 $\pm$  19.25\\
Entities per product (M\&R) & 181.11 $\pm$ 797.55 & 85.99 $\pm$ 276.30\\
\hline
\end{tabular}
\end{table}

\subsubsection{Evaluation Measures.} To quantify the quality of algorithms, we use Mean Reciprocal Rank (MRR), and average Recall@k (k= 5) and Normalized Discounted Cumulative Gain (NDCG) as the evaluation measures. We evaluate the performance of the algorithms for each individual user purchase observed in the data. Hence, for the same query but two different users, the relevant product (i.e. the product purchased) can be different. Therefore, in our dataset there is only a single relevant product for each query that resulted in a purchase, hence the use of MRR. Recall at rank 5, expresses whether the relevant document appears in the top-5 ranked products, while NDCG penalizes the effectiveness score by the rank at which the relevant product appears in a smoother way compared to MRR. %

\subsubsection{User Simulation.}
\label{UserSimu}
 Our experimentation depends on users responding to questions asked by our method. Conducting a user study is in our future plans, however, in this paper we follow recent work that simulates users ~\cite{zhang2018towards, sun2018conversational}.
We simulate users following two different settings: (1) we assume that the user will respond to the questions with full knowledge of whether an entity is present or not in the target product. Hence, we assume that a user has a product in mind, which is deterministic but unknown, and the user will respond with ``yes'' if an entity is contained in the target product documents and ``no'' if an entity is absent from the target product documents on the offline training phase and the online interactive search phase. This setting is the same used by ~\citet{zhang2018towards}, which assumes that the user has perfect knowledge of the value of an aspect for the target product; (2) additionally, we allow the users to give the wrong answer to our product search system with a given probability during online interactive search. We consider two noisy answers settings, which are described in Section ~\ref{sec:QSBPS}, while the precise experiment is described in \RQRef{4}.

\subsubsection{Baselines.} We compare our method with six baselines, in which the first three baselines are interactive baselines while the last three ones are query-product semantic matching baselines: (1) \textbf{Random}, which randomly chooses the entity from the entity pool to ask a question about; (2) \textbf{SBS} ~\cite{Wen:2013}, which is the sequential Bayesian search algorithm, that uses different training than our algorithm;
(3) \textbf{PMMN} ~\cite{zhang2018towards}, which is a state-of-the-art conversational recommender system asking questions on aspect-value pairs. Similar to the experiments run in the original paper, we assume that the system is able to retrieve the right candidate aspects for the product with 100\% accuracy, which leads to an upper bound performance impossible to be actually reached by a real system;
(4) \textbf{LSE} ~\cite{van2016learning}, which is one of the state-of-the-art latent vector models, that jointly learns the representations of words, products and the relationship between them in product search; 
(5) \textbf{TranSearch}
 ~\cite{guo2018multi}, which is one of the state-of-the-art product search models using multi-modal preference modeling from both textual and visual modalities. For fair consideration, we use the textual version of their TranSearch model, i.e., $\text{TranSearch}_t$ with pre-training; and (6) \textbf{ALSTP}
 ~\cite{guo2019attentive}, which is one of the state-of-the-art product search models using attention networks of long-term and short-term user preferences.
For the latter four baselines, we use the optimal parameters reported in the corresponding product search papers. 

\begin{figure}[tb]
\captionsetup{font={small}}
  \caption{Heatmap of the MRR results on the validation set of "Home \& Kitchen" (top), and "Clothing, Shoes \& Jewelry" (bottom). The MRR is shown as a function of the number of questions asked and the weight of question reward $\gamma$. The more red the heatmap, the better the performance of the method. The optimal weight parameter $\gamma$ for the corresponding number of questions asked is designated by the white boundary box and is reported in the table below the heatmap. (Unless mentioned otherwise, in what follows the optimal weight parameter $\gamma$ we used for the corresponding number of questions asked is designated in the figure).
}
  \includegraphics[width=0.75\columnwidth]{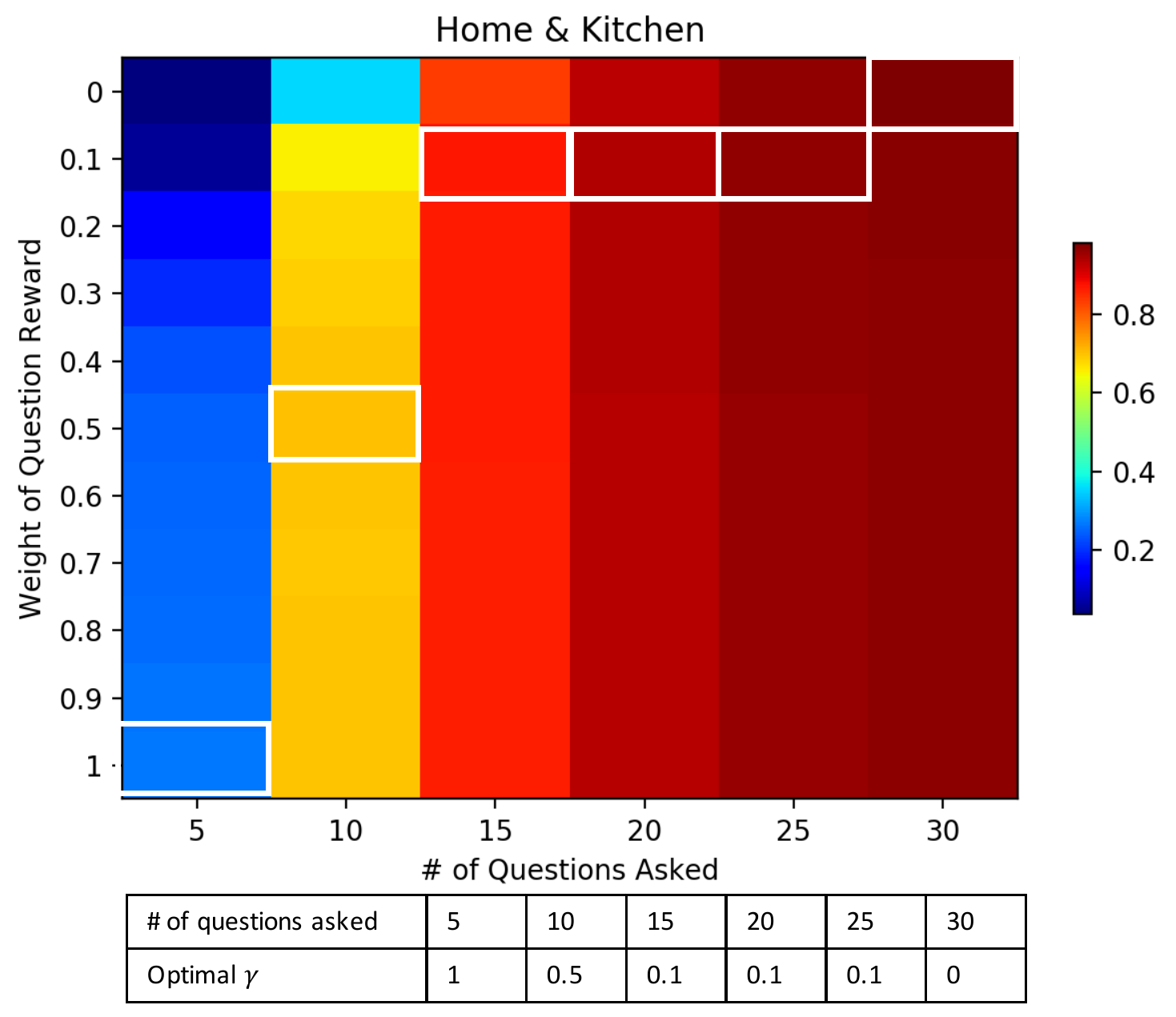}
  \includegraphics[width=0.75\columnwidth]{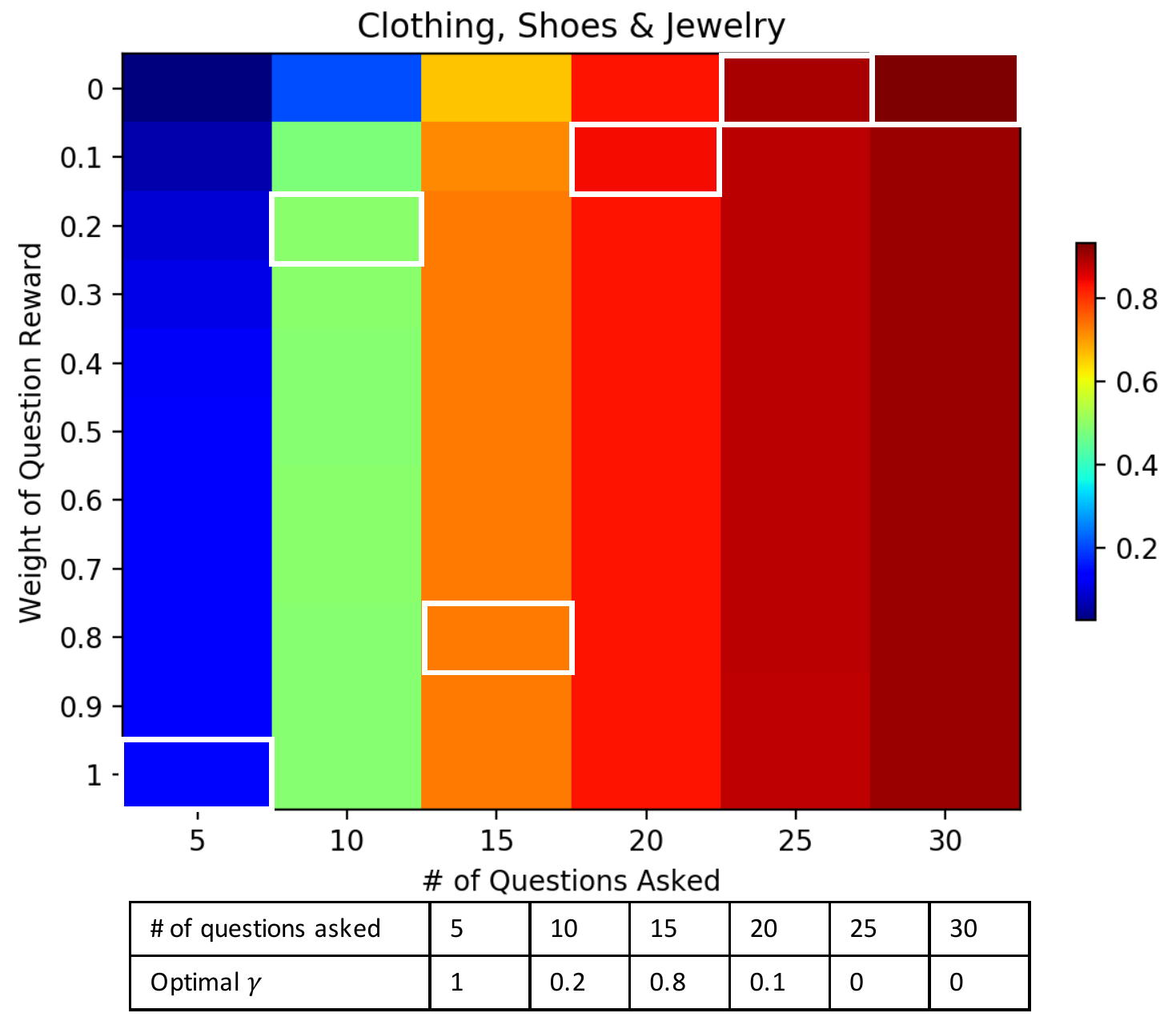}
  \label{fig:heatmap}
\end{figure}

\begin{table*}
\captionsetup{font={small}}
\caption{The comparison of MRR, and Recall$@$5, and NDCG results on "Home \& Kitchen" (top), and "Clothing, Shoes \& Jewelry" (bottom) by using metadata only versus using metadata \& reviews. As it can be observed, user reviews, while noisy, can improve the selection of informative questions, by discussing interesting properties/entities of the products not present in the product descriptions.}
\label{table:3}
\centering
\small
\begin{tabular}{c|c|c|c|c|c|c}
\hline
 & \multicolumn{2}{p{0.4\columnwidth}<{\centering}|}{MRR} & \multicolumn{2}{p{0.4\columnwidth}<{\centering}|}{Recall$@$5} & \multicolumn{2}{p{0.4\columnwidth}<{\centering}}{NDCG} \\
\hline 
\# of questions asked & Meta only & Meta \& review & Meta only & Meta \& review & Meta only & Meta \& review \\
\hline 
5 & 0.290 & 0.292 & 0.427 & 0.439 & 0.411 & 0.423\\
10 & 0.568 & 0.684 & 0.690 & 0.809 & 0.647 & 0.749\\
15 & 0.702 & 0.860 & 0.790 & 0.923 & 0.758 & 0.890\\
20 & 0.779 & 0.932 & 0.855 & 0.965 & 0.821 & 0.947\\
25 & 0.822 & 0.956 & 0.890 & 0.977 & 0.856 & 0.966\\
30 & 0.846 & 0.982 & 0.862 & 0.984 & 0.870 & 0.985\\
\hline
\end{tabular}
\begin{tabular}{c|c|c|c|c|c|c}
\hline
 & \multicolumn{2}{p{0.4\columnwidth}<{\centering}|}{MRR} & \multicolumn{2}{p{0.4\columnwidth}<{\centering}|}{Recall$@$5} & \multicolumn{2}{p{0.4\columnwidth}<{\centering}}{NDCG} \\
\hline 
\# of questions asked & Meta only & Meta \& review & Meta only & Meta \& review & Meta only & Meta \& review \\
\hline
 5 & 0.140 & 0.145 & 0.213 & 0.213 & 0.274 & 0.288\\
10 & 0.327 & 0.486 & 0.462 & 0.645 & 0.445 & 0.588\\
15 & 0.490 & 0.719 & 0.630 & 0.835 & 0.585 & 0.777\\
20 & 0.575 & 0.819 & 0.715 & 0.906 & 0.657 & 0.859\\
25 & 0.668 & 0.897 & 0.736 & 0.928 & 0.724 & 0.917\\
30 & 0.745 & 0.930 & 0.802 & 0.960 & 0.789 & 0.945\\
\hline
\end{tabular}
\end{table*}

\begin{table*}
\captionsetup{font={small}}
\caption{The performance of our algorithm when excluding all training, when including only questions effectiveness training, when including only product relevance training, and performing the proposed duet training by using other users' data, represented in the table by No-train, Q-train, P-train, and Duet, respectively on the "Home \& Kitchen" (top) and  "Clothing, Shoes \& Jewelry" (bottom). As it can be observed Duet training outperforms all other options demonstrating the suitability of the proposed method.}
\label{table:4}
\centering
 \small
\begin{tabular}{c|cccc|cccc|cccc}
\hline
 & \multicolumn{4}{p{0.4\columnwidth}<{\centering}|}{MRR} & \multicolumn{4}{p{0.4\columnwidth}<{\centering}|}{Recall $@$ 5} & \multicolumn{4}{p{0.4\columnwidth}<{\centering}}{NDCG}\\
\hline 
\# of questions asked & No-train & Q-train & P-train & Duet & No-train & Q-train & P-train & Duet & No-train & Q-train & P-train & Duet\\
\hline 
5 & 0.00 & 0.14 & 0.04 & 0.29 & 0 & 0.21 & 0.05 & 0.44 & 0.12 & 0.29 & 0.18 & 0.42\\
10 & 0.13 & 0.53 & 0.35 & 0.68 & 0.27 & 0.69 & 0.59 & 0.81 & 0.30 & 0.63 & 0.49 & 0.75\\
15 & 0.78 & 0.77 & 0.84 & 0.86 & 0.82 & 0.85 & 0.89 & 0.92 & 0.82 & 0.81 & 0.87 & 0.89\\
20 & 0.91 & 0.87 & 0.93 & 0.93 & 0.92 & 0.92 & 0.95 & 0.97 & 0.93 & 0.89 & 0.95 & 0.95\\
25 & 0.95 & 0.91 & 0.96 & 0.96 & 0.96 & 0.94 & 0.97 & 0.98 & 0.96 & 0.93 & 0.97 & 0.97\\
30 & 0.98 & 0.92 & 0.98 & 0.98 & 0.98 & 0.95 & 0.98 & 0.98 & 0.98 & 0.94 & 0.99 & 0.99\\
\hline
\end{tabular}
\begin{tabular}{c|cccc|cccc|cccc}
\hline
 & \multicolumn{4}{p{0.4\columnwidth}<{\centering}|}{MRR} & \multicolumn{4}{p{0.4\columnwidth}<{\centering}|}{Recall $@$ 5} & \multicolumn{4}{p{0.4\columnwidth}<{\centering}}{NDCG}\\
\hline 
\# of questions asked & No-train & Q-train & P-train & Duet & No-train & Q-train & P-train & Duet & No-train & Q-train & P-train & Duet\\
\hline 
5 & 0.00 & 0.05 & 0.03 & 0.15 & 0 & 0.07 & 0.04 & 0.21 & 0.11 & 0.19 & 0.16 & 0.29\\
10 & 0.06 & 0.33 & 0.21 & 0.49 & 0.07 & 0.48 & 0.34 & 0.65 & 0.21 & 0.45 & 0.35 & 0.59\\
15 & 0.57 & 0.60 & 0.66 & 0.72 & 0.65 & 0.73 & 0.74 & 0.84 & 0.65 & 0.68 & 0.72 & 0.78\\
20 & 0.79 & 0.73 & 0.82 & 0.82 & 0.83 & 0.83 & 0.88 & 0.91 & 0.83 & 0.79 & 0.86 & 0.86\\
25 & 0.88 & 0.81 & 0.90 & 0.90 & 0.90 & 0.89 & 0.93 & 0.93 & 0.90 & 0.85 & 0.92 & 0.92\\
30 & 0.92 & 0.85 & 0.93 & 0.93 & 0.94 & 0.92 & 0.96 & 0.96 & 0.93 & 0.88 & 0.94 & 0.94\\
\hline
\end{tabular}
\end{table*}

\subsection{The Impact of the Number of Questions and the Question Reward Parameter $\gamma$}

In this section we answer \RQRef{1}. Our proposed method is parameterized by the number of questions to be asked to the user and the hyper-parameter $\gamma$ which is the weight of question reward in QSBPS. We evaluate the results on the validation set. The number of question asked to the user ranges from 5 to 30 with an interval of 5, and the weight $\gamma$ ranges from 0 to 1 with an interval of 0.1. Figure ~\ref{fig:heatmap} shows the heat map of MRR results for every combination of the number of asked questions and the question training controlling parameter $\gamma$ on "Home \& Kitchen" and "Clothing, Shoes \& Jewelry" categories. The x-axis is the number of questions asked, and the y-axis is different value of weight parameter $\gamma$. From Figure ~\ref{fig:heatmap} it can be seen that the MRR increases with the number of questions asked to the user, as expected: the more the questions asked the better the performance of the algorithm. It can also be observed that MRR fluctuates over different question reward controlling parameter $\gamma$. Different number of asked questions produce different optimal value for $\gamma$. The optimal $\gamma$ for 5, 10, 15, 20, 25, and 30 questions is 1, 0.5, 0.1, 0.1, 0.1, and 0, respectively on "Home \& Kitchen", while 1, 0.2, 0.8, 0.1, 0, and 0, respectively, on "Clothing, Shoes \& Jewelry". As we can see, the overall trend of optimal $\gamma$ is decreasing with the number of questions. This suggest that question reward training is more important in the early stages of question asking, especially when the number of questions is 5 and 10. When the number of questions asked is large, the importance of question reward training decreases. The reason could be that a large number of questions are sufficient for high performance regardless of reward.

\subsection{The Influence of Using User Reviews Data}
\label{RQ4.3}

For \RQRef{2}, we explore the impact of using the user reviews data. We compare the differences between the results using products meta data only and the results using both products meta data and user reviews data. The comparison results are shown in Table ~\ref{table:3} with three metrics, MRR, Recall$@$5 and NDCG on the two categories. For each number of questions asked, the near-optimal $\gamma$ was used as indicated by the white-boundary boxes and tables in Figure ~\ref{fig:heatmap}. As shown in Table ~\ref{table:3}, we can see that three metrics are higher for the combined meta data and reviews data compared to only using meta data, especially when the number of questions is greater than 5. This indicates that user reviews are important in the users buying process, and offer entities that can be more discriminative than the ones included in the product description documents. %

\begin{table*}
\captionsetup{font={small}}
\caption{The results of noisy answers on "Home \& Kitchen" (top), and "Clothing, Shoes \& Jewelry" (bottom)  when $h(e)$ is a fixed, ranging from10\% to 50\% with a step 10\%. Our method outperforms the best query-product matching baselines, the TranSearch and ALSTP model, which are presented in Table~\ref{table:baseline}, after about 3 questions asked with 10\% of wrong answers, and after about 6 questions being asked with 20\% of wrong answers. Naturally the performance of our method is equal to random ranking when wrong answers are provided 50\% of the time.}
\label{table:noisy}
\centering
  \small
\begin{tabular}{p{0.2\columnwidth}<{\centering}|p{0.06\columnwidth}<{\centering}p{0.06\columnwidth}<{\centering}p{0.06\columnwidth}<{\centering}p{0.06\columnwidth}<{\centering}p{0.06\columnwidth}<{\centering}p{0.06\columnwidth}<{\centering}|p{0.06\columnwidth}<{\centering}p{0.06\columnwidth}<{\centering}p{0.06\columnwidth}<{\centering}p{0.06\columnwidth}<{\centering}p{0.06\columnwidth}<{\centering}p{0.06\columnwidth}<{\centering}|p{0.06\columnwidth}<{\centering}p{0.06\columnwidth}<{\centering}p{0.06\columnwidth}<{\centering}p{0.06\columnwidth}<{\centering}p{0.06\columnwidth}<{\centering}p{0.06\columnwidth}<{\centering}}
\hline
 & \multicolumn{6}{p{0.4\columnwidth}<{\centering}|}{MRR} & \multicolumn{6}{p{0.4\columnwidth}<{\centering}|}{Recall $@$ 5} & \multicolumn{6}{p{0.4\columnwidth}<{\centering}}{NDCG}\\
\hline 
\small{\# of questions asked} & No-noise  & 10\% & 20\% & 30\% & 40\% & 50\%  & No-noise  & 10\% & 20\% & 30\% & 40\% & 50\%  & No-noise & 10\% & 20\% & 30\% & 40\% & 50\% \\
\hline 
5 & 0.292 & 0.186 & 0.111 & 0.063 & 0.033 & 0.016 & 0.439 & 0.274 & 0.166 & 0.088 & 0.049 & 0.024 & 0.423 & 0.313 & 0.232 & 0.175 & 0.135 & 0.111\\
10 & 0.684 & 0.398 & 0.194 & 0.082 & 0.033 & 0.012 & 0.809 & 0.507 & 0.268 & 0.117 & 0.045 & 0.016 & 0.749 & 0.501 & 0.312 & 0.196 & 0.138 & 0.107\\
15 & 0.860 & 0.538 & 0.283 & 0.114 & 0.033 & 0.007 & 0.923 & 0.640 & 0.373 & 0.152 & 0.044 & 0.008 & 0.890 & 0.622 & 0.396 & 0.229 & 0.138 & 0.100\\
20 & 0.932 & 0.651 & 0.342 & 0.130 & 0.036 & 0.007 & 0.965 & 0.752 & 0.433 & 0.169 & 0.050 & 0.010 & 0.947 & 0.718 & 0.450 & 0.247 & 0.142 & 0.099\\
\hline
\end{tabular}
\begin{tabular}{p{0.2\columnwidth}<{\centering}|p{0.06\columnwidth}<{\centering}p{0.06\columnwidth}<{\centering}p{0.06\columnwidth}<{\centering}p{0.06\columnwidth}<{\centering}p{0.06\columnwidth}<{\centering}p{0.06\columnwidth}<{\centering}|p{0.06\columnwidth}<{\centering}p{0.06\columnwidth}<{\centering}p{0.06\columnwidth}<{\centering}p{0.06\columnwidth}<{\centering}p{0.06\columnwidth}<{\centering}p{0.06\columnwidth}<{\centering}|p{0.06\columnwidth}<{\centering}p{0.06\columnwidth}<{\centering}p{0.06\columnwidth}<{\centering}p{0.06\columnwidth}<{\centering}p{0.06\columnwidth}<{\centering}p{0.06\columnwidth}<{\centering}}
\hline
 & \multicolumn{6}{p{0.4\columnwidth}<{\centering}|}{MRR} & \multicolumn{6}{p{0.4\columnwidth}<{\centering}|}{Recall $@$ 5} & \multicolumn{6}{p{0.4\columnwidth}<{\centering}}{NDCG}\\
\hline 
\small{\# of questions asked} & No-noise  & 10\% & 20\% & 30\% & 40\% & 50\%  & No-noise  & 10\% & 20\% & 30\% & 40\% & 50\%  & No-noise & 10\% & 20\% & 30\% & 40\% & 50\% \\
\hline 
5 & 0.145 & 0.090 & 0.052 & 0.029 & 0.016 & 0.008 & 0.213 & 0.129 & 0.074 & 0.040 & 0.022 & 0.010 & 0.288 & 0.219 & 0.168 & 0.135 & 0.110 & 0.095\\
10 & 0.486 & 0.238 & 0.110 & 0.047 & 0.018 & 0.007 & 0.645 & 0.325 & 0.157 & 0.068 & 0.023 & 0.007 & 0.588 & 0.361 & 0.230 & 0.156 & 0.113 & 0.092\\
15 & 0.719 & 0.393 & 0.175 & 0.064 & 0.019 & 0.005 & 0.835 & 0.507 & 0.231 & 0.088 & 0.025 & 0.007 & 0.777 & 0.498 & 0.294 & 0.176 & 0.116 & 0.090\\
20 & 0.819 & 0.505 & 0.230 & 0.082 & 0.018 & 0.004 & 0.906 & 0.616 & 0.312 & 0.114 & 0.025 & 0.005 & 0.859 & 0.595 & 0.349 & 0.194 & 0.116 & 0.089\\
\hline
\end{tabular}
\end{table*}

\begin{table*}
\captionsetup{font={small}}
\caption{The results of noisy answers on "Home \& Kitchen" (top), and "Clothing, Shoes \& Jewelry" (bottom)  when $h(e)$ is modelled by term frequency. Our method, when the optimal $\beta$ is being used, outperforms the best query-product matching baselines, presented in Table~\ref{table:baseline}, after about 4 questions are being asked, despite the noise in the answers.
}
\label{table:noisy_3}
\centering
  \small
\begin{tabular}{c|ccc|ccc|ccc}
\hline
 & \multicolumn{3}{p{0.4\columnwidth}<{\centering}|}{MRR} & \multicolumn{3}{p{0.4\columnwidth}<{\centering}|}{Recall $@$ 5} & \multicolumn{3}{p{0.4\columnwidth}<{\centering}}{NDCG}\\
\hline 
\# of questions asked & No-noise & optimal $\beta$ & $\beta$=0 & No-noise & optimal $\beta$ & $\beta$=0 & No-noise & optimal $\beta$ & $\beta$=0 \\
\hline
5 & 0.292 & 0.156 & 0.129 & 0.439 & 0.232 & 0.185 & 0.423 & 0.281 & 0.248\\
10 & 0.684 & 0.245 & 0.194 & 0.809 & 0.325 & 0.264 & 0.749 & 0.358 & 0.308\\
15 & 0.860 & 0.295 & 0.220 & 0.923 & 0.376 & 0.283 & 0.890 & 0.401 & 0.329\\
20 & 0.932 & 0.330 & 0.250 & 0.965 & 0.400 & 0.308 & 0.947 & 0.431 & 0.356\\
\hline
\end{tabular}
\begin{tabular}{c|ccc|ccc|ccc}
\hline
 & \multicolumn{3}{p{0.4\columnwidth}<{\centering}|}{MRR} & \multicolumn{3}{p{0.4\columnwidth}<{\centering}|}{Recall $@$ 5} & \multicolumn{3}{p{0.4\columnwidth}<{\centering}}{NDCG}\\
\hline 
\# of questions asked & No-noise & optimal $\beta$ & $\beta$=0 & No-noise & optimal $\beta$ & $\beta$=0 & No-noise & optimal $\beta$ & $\beta$=0 \\
\hline 
5 & 0.145 & 0.061 & 0.048 & 0.213 & 0.090 & 0.069 & 0.288 & 0.176 & 0.160\\
10 & 0.486 & 0.095 & 0.068 & 0.645 & 0.133 & 0.094 & 0.588 & 0.209 & 0.181\\
15 & 0.719 & 0.112 & 0.091 & 0.835 & 0.152 & 0.123 & 0.777 & 0.224 & 0.204\\
20 & 0.819 & 0.126 & 0.106 & 0.906 & 0.170 & 0.142 & 0.859 & 0.239 & 0.217\\
\hline
\end{tabular}
\end{table*}

\subsection{The Influence of the Duet Training}
To answer \RQRef{3}, we investigate the impact of our duet training using other users' data, i.e. learning both the questions performance over entity and the system belief over products, by comparing it to (1) using no training, (2) using only questions performance training on entities, and (3) using only system belief training on products. Our hypothesis is that the historical data of a specific user together with that of other users captures important information and the model training of our duet learning framework from these data will be beneficial even if the training data does not exactly match the information of the specific target user. Indeed, Table ~\ref{table:4} shows three metrics using duet training are higher than the ones that corresponds to excluding one questions training, system belief training or both, especially when the number of questions is less than 20. For each number of questions asked, the near-optimal $\gamma$ was used as indicated by the white-boundary boxes and tables in Figure ~\ref{fig:heatmap}. 
We conclude that using our duet learning framework is highly beneficial. When large number of questions (greater than 15) are asked, the "No-train" achieve higher performance than "Q-train". This might be because "Q-train" only use the noisy rewards trained by the weak signals and do not use GBS policy like "No-train","P-train", and "Duet" to select the optimal question to ask, and thus less effective when the number of questions are getting high.

\begin{table*}
\captionsetup{font={small}}
\caption{NDCG and Recall$@$5 on "Home \& Kitchen" (top), and "Clothing, Shoes \& Jewelry" (bottom) for the compared methods. \#. represents the number of questions asked. }
\label{table:baseline}
\centering
  \small
\begin{tabular}{c|ccccccc|ccccccc}
\hline
 & \multicolumn{7}{p{0.9\columnwidth}<{\centering}|}{NDCG} & \multicolumn{7}{p{0.9\columnwidth}<{\centering}}{Recall $@$ 5} \\
\hline
\#. & Random & LSE & TranSearch & ALSTP & PMMN & SBS & QSBPS & Random & LSE & TranSearch & ALSTP & PMMN & SBS & QSBPS\\
\hline
5 & 0.011 & 0.176 & 0.198 & 0.181 & 0.271 & 0.001 & \textbf{0.401} & 0.006 & 0.074 & 0.131 & 0.149 & 0.158 & 0 & \textbf{0.439}\\
10 & 0.012 & 0.176 & 0.198 & 0.181 & 0,285 & 0.267 & \textbf{0.742} & 0.006 & 0.074 & 0.131 & 0.149 & 0.188 & 0.267 & \textbf{0.809}\\
15 & 0.015 & 0.176 & 0.198 & 0.181 & 0.286 & 0.815 & \textbf{0.888} & 0.010 & 0.074 & 0.131 & 0.149 & 0.188 & 0.828 & \textbf{0.923}\\
20 & 0.014 & 0.176 & 0.198 & 0.181 & 0.286 & 0.923 & \textbf{0.946} & 0.009 & 0.074 & 0.131 & 0.149 & 0.188 & 0.919 & \textbf{0.965}\\
25 & 0.015 & 0.176 & 0.198 & 0.181 & 0.286 & 0.960 & \textbf{0.966} & 0.008 & 0.074 & 0.131 & 0.149 & 0.188 & 0.961 & \textbf{0.977}\\
30 & 0.016 & 0.176 & 0.198 & 0.181 & 0.286 & 0.980 & \textbf{0.985} & 0.011 & 0.074 & 0.131 & 0.149 & 0.188 & 0.981 & \textbf{0.984}\\
\hline
\end{tabular}
\begin{tabular}{c|ccccccc|ccccccc}
\hline
 & \multicolumn{7}{p{0.9\columnwidth}<{\centering}|}{NDCG} & \multicolumn{7}{p{0.9\columnwidth}<{\centering}}{Recall $@$ 5} \\
\hline
\#. & Random & LSE & TranSearch & ALSTP & PMMN & SBS & QSBPS & Random & LSE & TranSearch & ALSTP & PMMN & SBS & QSBPS\\
\hline
 5 & 0 & 0.098 & 0.101 & 0.131 & 0.231 & 0.001 & \textbf{0.256} & 0 & 0.045 & 0.056 & 0.083 &  0.101 & 0 & \textbf{0.213}\\
10 & 0 & 0.098 & 0.101 & 0.131 & 0.248 & 0.178 & \textbf{0.577} & 0 & 0.045 & 0.056 & 0.083 & 0.13 &  0.075 & \textbf{0.645}\\
15 & 0.001 & 0.098 & 0.101 & 0.131 & 0.249 & 0.632 & \textbf{0.772} & 0.001 & 0.045 & 0.056 & 0.083 & 0.13 & 0.664 & \textbf{0.835}\\
20 & 0.001 & 0.098 & 0.101 & 0.131 & 0.249 & 0.817 & \textbf{0.856} & 0.001 & 0.045 & 0.056 & 0.083 & 0.13 & 0.833 & \textbf{0.906}\\
25 & 0.001 & 0.098 & 0.101 & 0.131 & 0.249 & 0.895 & \textbf{0.915} & 0.002 & 0.045 & 0.056 & 0.083 & 0.13 & 0.904 & \textbf{0.928}\\
30 & 0.001 & 0.098 & 0.101 & 0.131 & 0.249 & 0.931 & \textbf{0.943} & 0.001 & 0.045 & 0.056 & 0.083 & 0.13 & 0.944 & \textbf{0.960}\\
\hline
\end{tabular}
\end{table*}

\subsection{The Influence of Noisy Answers}

Given that the user may not always give us the right answer, we also explore the noise-tolerance of our QSBPS algorithm towards answering \RQRef{4}. We develop a noise-tolerant version our QSBPS algorithm as shown in Section ~\ref{sec:QSBPS} and investigate what the influence of noisy answers is. We simulate the noisy answer of the user under two settings. In the first setting, we fix the probability of error, $\varepsilon$, and consider it as a parameter that ranges from 10\% to 50\% with a step 10\%. The results when $h(e)$ is a fixed number $\varepsilon$ is shown in Table ~\ref{table:noisy}. For each number of questions asked, the near-optimal $\gamma$ was used as indicated by the white-boundary boxes and tables in Figure ~\ref{fig:heatmap}. It is obvious and expected that the performance as captured by three metrics decreases with the increase of $\varepsilon$. When $\varepsilon$ is equal to 50\%, which means the user answer the question with "yes" or "no" at random, we observe very low performance. On the other hand, when $\varepsilon$ is equal to 10\%, the values of three metrics are still relatively high. Further, note that, in comparison to the best query-product matching baselines, the TranSearch and ALSTP model, which is presented in Table~\ref{table:baseline}, for 10\% wrong answers, our method outperforms the baseline after 3 questions asked, and for 20\% wrong answers, after 6 questions being asked.

In the second setup, we define $h(e)$ as a function of term frequency of entity as shown in Equation ~\ref{equ:errorrate}. Similar to \RQRef{1}, we evaluate the MRR on validation set for the two categories, to select the optimal $\beta$. %
Due to space limitations, the heatmaps were omitted and we only report the optimal $\beta$ here. 
The optimal $\beta$ is $1$ on "Home \& Kitchen", and $0.8$ on "Clothing, Shoes \& Jewelry". The results when $h(e)$ is modelled by term frequency using the optimal weight parameter $\beta$ are shown in Table ~\ref{table:noisy_3}. For each number of questions asked, the near-optimal $\gamma$ was used as indicated by the white-boundary boxes and tables in Figure ~\ref{fig:heatmap}. As one can observe, the performance as measured by the three metrics when using the optimal $\beta$ is higher than that when $\beta$=0, which suggests that the objective function of our noise-tolerant version of QSBPS algorithm is effective. Further, note that our method, when the optimal $\beta$ is being used, outperforms the best query-product matching baselines, the TranSearch and ALSTP model, presented in Table~\ref{table:baseline}, after 4 questions are being asked, despite the noise in the answers.

\subsection{The Effectiveness of Our Proposed Method Compared with Other Algorithms}

In \RQRef{5}, we answer how effective is our proposed method for finding the best matching product compared to the six baselines shown in Section ~\ref{ExpSetup}, by reporting NDCG and Recall$@$5. 
Table ~\ref{table:baseline} shows the results. For our method, for each number of questions asked, the near-optimal $\gamma$ was used as indicated by the white-boundary boxes and tables in Figure ~\ref{fig:heatmap}. As indicated in Table ~\ref{table:baseline}, our QSBPS algorithm achieves the highest effectiveness scores when compared to Random, LSE, SBS, TranSearch, ALSTP, and PMMN. Our QSBPS algorithm exceeds Random, which indicates, as expected that our question selection strategy is better than choosing questions in random. After less than 5 questions, our QSBPS algorithm greatly improve over the LSE, TranSearch, and ALSTP model. This clearly suggests that a theoretically optimal sequence of entity-centered questions can be rather helpful and greatly improve the performance in product search. Our QSBPS algorithm perform better than SBS, which indicates the effectiveness of our cross-user duet training.
Last, our QSBPS model proves to be better than the interactive PMMN system. This can be explained, by either the fact that our pool of questions is better, or the fact that our question sequence strategy is better. Given that in both systems, QSBPS, the question strategy is strongly connected to the type of questions placed in the question pool, it makes it very hard to decompose the effect in the improvements demonstrated by QSBPS.
One final observation on PMMN is that its performance almost does not increase when the number of questions is larger than 10. The reason for this is the fact that it is rather difficult to extract more than 10 aspect-value pairs from each user review for a certain item. As a consequence, there are no more available questions to ask.

\section{Conclusions and Discussion}
\label{sec:conc}
The focus of this work is helping users to find the most relevant product in a large repository. We propose a question-based sequential Bayesian product search algorithm, called QSBPS, which efficiently locates the most relevant product by directly querying users on the presence or absence of an informative term in product related documents.
Our framework first identifies and extracts informative terms (instantiated by entities in this work) mentioned in the text of the given product, and then trains a system belief and question reward model by using historical data. Based on the trained model, our method derives a policy of the optimal questions to be asked to the user. After receiving an answer to each question asked to the user, the posterior product preference of the user is calculated to generate the ranked recommendation list. Experiments on the Amazon product dataset demonstrate the effectiveness of QSBPS compared to state-of-the-art. Further, we illustrate the performance of our method when noisy answers are received by users.

\label{sec:discuss}
In this work we pivot around the presence or absence of entities in the target products to generate a pool of questions to be asked to the user, which is still a rudimentary method of generating questions. Clearly, there is a richer set of possible questions to be asked, questions that may or may not be answered by a ``yes'' or a ``no''. Questions similar to the ones we have constructed could also be constructed by using labelled topics ~\cite{ZOU201719}, keywords extraction ~\cite{campos2018text}, item categories and attributes, or extracted triplets ~\cite{reddy2017generating}. Richer type of questions could be also constructed by identifying properties of the products in the descriptions and reviews and their relation to the product. For example, for a ``Canon EOS 5D Mark II'' digital camera, the following relations could be identified in the product description: ``resolution'', ``manufacturer'', ``LCD screen dimension''.
We leave this as future work.
Further, our works simulates user answers, noisy or not. A user study can be particularly helpful in understanding whether users are willing to answer a small number of questions, under what conditions (e.g., they may be willing if they have already reformulated their query a number of times), and to what extent they can provide correct answers.
From a technical perspective, our work proposes a stand-alone algorithm that learns the informativeness of questions, along with user preferences. In principle, however, one can use a ranking method (any of the baselines) to construct an informative prior belief on user preferences and reduce the number of necessary questions to find the product to smaller than 5. Further, one can also incorporate other factors (e.g, the importance level of different informative terms) to the objective function of question selection to extend the work.
Furthermore, in this work we made the assumption that we know the topic of a user's query, so that we can load the right prior over preferences, and entity rewards. In practice, one needs some technique (of text similarity) to soft-match an arbitrary query to the already known, which we intent to explore in the future.
Last, it is highly likely that an entity is semantically related to the desired product but it is not lexically contained in the description of it. In this work we do not explore any semantic correlation modeling, but we leave it as the future work. 

\bibliographystyle{ACM-Reference-Format}
\bibliography{bibfile} 

\end{document}